\documentstyle[12pt,epsfig,graphics]{article}
\begin{document}\bibliographystyle{plain}\begin{titlepage}
\renewcommand{\thefootnote}{\fnsymbol{footnote}}\hfill\begin{tabular}{l}
HEPHY-PUB 731/00\\UWThPh-2000-30\\hep-ph/0009185\\August 2000\end{tabular}
\\[1cm]\Large\begin{center}{\bf INSTANTANEOUS BETHE--SALPETER EQUATION:
UTMOST ANALYTIC APPROACH}\\\vspace{0.8cm}\large{\bf Wolfgang
LUCHA\footnote[1]{\normalsize\ {\em E-mail address\/}:
wolfgang.lucha@oeaw.ac.at}}\\[.3cm]\normalsize Institut f\"ur
Hochenergiephysik,\\\"Osterreichische Akademie der
Wissenschaften,\\Nikolsdorfergasse 18, A-1050 Wien,
Austria\\[0.7cm]\large{\bf Khin MAUNG MAUNG\footnote[3]{\normalsize\ {\em
E-mail address\/}: maung@jlab.org}}\\[.3cm]\normalsize Department of Physics,
Hampton University,\\Hampton, VA 23668\\[0.7cm]\large{\bf Franz F.
SCH\"OBERL\footnote[2]{\normalsize\ {\em E-mail address\/}:
franz.schoeberl@univie.ac.at}}\\[.3cm]\normalsize Institut f\"ur Theoretische
Physik, Universit\"at Wien,\\Boltzmanngasse 5, A-1090 Wien,
Austria\vfill{\normalsize\bf Abstract}\end{center}\normalsize The
Bethe--Salpeter formalism in the instantaneous approximation for the
interaction kernel entering into the Bethe--Salpeter equation represents a
reasonable framework~for the description of bound states within relativistic
quantum field theory. In contrast~to its further simplifications (like, for
instance, the so-called reduced Salpeter equation), it allows also the
consideration of bound states composed of ``light'' constituents.~Every
eigenvalue equation with solutions in some linear space may be
(approximately) solved by conversion into an equivalent matrix eigenvalue
problem. We demonstrate that~the matrices arising in these representations of
the instantaneous Bethe--Salpeter equation may be found, at least for a wide
class of interactions, in an entirely algebraic manner. The advantages of
having the involved matrices explicitly, i.e., not ``contaminated''~by errors
induced by numerical computations, at one's disposal are obvious: problems
like, for instance, questions of the stability of eigenvalues may be analyzed
more rigorously; furthermore, for small matrix sizes the eigenvalues may even
be calculated analytically.\vspace{3ex}

\renewcommand{\thefootnote}{\arabic{footnote}}\end{titlepage}

\normalsize

\section{Introduction}The Bethe--Salpeter formalism \cite{BSE,SE} is
generally accepted to represent the appropriate framework for the description
of bound states within relativistic quantum field theory. Within this
formalism, a bound state is described by its ``Bethe--Salpeter amplitude,''
which is defined as the (time-ordered) product of the field operators of the
bound-state constituents between the vacuum and the bound state. In
principle, this bound-state amplitude should be found as solution of the
(homogeneous) Bethe--Salpeter equation.

However, apart from very few special cases---like the famous Wick--Cutkosky
model which describes the interaction of two scalar particles by exchange of
a massless scalar particle---the Bethe--Salpeter equation turns out to be
practically not tractable. One~of the main reasons for this fact is the
appearance of timelike variables in the equation~of motion. Consequently
people usually consider some three-dimensional reduction~of~the
Bethe--Salpeter equation. The most popular among these three-dimensional
reductions is based on the assumption that the interaction between the
bound-state constituents is instantaneous in the center-of-momentum frame of
the bound state; the result of~this is called the ``Salpeter equation'' or
``instantaneous Bethe--Salpeter equation'' \cite{SE}. This equation may be
formulated as eigenvalue problem for the mass $M$ of the bound state. Its
dynamical quantity is the ``Salpeter amplitude,'' obtained from the
Bethe--Salpeter amplitude by equating the time variables of the involved
bound-state constituents.

By expanding the Salpeter amplitude into some convenient set of basis
matrices in Dirac space \cite{leyaouanc85}, Laga\"e \cite{lagae92I,lagae92II}
managed to reduce the full instantaneous Bethe--Salpeter equation for
fermion--antifermion bound states to a set of coupled equations for radial
wave functions. His easy-to-handle formalism immediately stimulated renewed
interest in the full Salpeter equation as a manageable tool for describing
mesonic bound states
\cite{resag94,muenz94,parramore95,parramore96,Zoeller95,olsson95}. The first
question to be answered in this context refers to the Lorentz structure of
the confining Bethe--Salpeter kernel. The most important outcome of these
studies is the following:\begin{itemize}\item Lorentz-scalar confinement
turns out (1) to face serious problems with respect~to the stability of the
resulting solutions \cite{parramore95,parramore96,olsson95} and (2) to be in
conflict with~the phenomenologically dictated linearity of the mesonic Regge
trajectories \cite{lagae92II,muenz94,olsson95}.\item Time-component
Lorentz-vector confinement, on the other hand, does not share either of these
two problems, that is, it yields (1) stable solutions
\cite{parramore95,parramore96,olsson95} and~(2) linear Regge trajectories for
the bound states of quark--antiquark systems
\cite{muenz94,olsson95}.\end{itemize}The square of the norm of the amplitudes
obtained as solutions of the instantaneous Bethe--Salpeter equation is not
positive definite; in order to stay in the physical sector of the theory, in
all these analyses one has to retain only bound states of positive~norm
squared but to reject all bound states of zero or negative norm squared. In
this sector, all mass eigenvalues $M$ of the instantaneous Bethe--Salpeter
equation are guaranteed to be real \cite{lagae92I,resag94}.

A standard technique for determining the solutions of a given eigenvalue
equation~is the conversion of this equation into the equivalent matrix
eigenvalue problem. (For the instantaneous Bethe--Salpeter equation, this
route has been followed in Refs.~\cite{lagae92I,resag94,olsson95}.) We show
here that, for a suitable choice of basis states in the space of solutions
\cite{Jacobs86}, the matrix representation of the instantaneous
Bethe--Salpeter equation may be explicitly given in analytical form. Among
others, this might prove to be advantageous for future discussions of the
stability of solutions of the instantaneous Bethe--Salpeter equation.

\section{The Instantaneous Bethe--Salpeter Equation}It goes without saying
that we would like to demonstrate our procedure for solving~the instantaneous
Bethe--Salpeter equation at the simplest conceivable example. First of all,
let us agree on two simplifying assumptions:\begin{itemize}\item The (full)
fermion propagators $S_i(p_i,m_i),$ $i=1,2,$ entering in the Bethe--Salpeter
equation may be approximated by the corresponding free propagators, given
by$$S_{i,0}^{-1}(p_i,m_i)=-{\rm i}\,(\gamma_\mu\,p_i^\mu-m_i)\ ,\quad i=1,2\
.$$In the above free-propagator approximation, the masses $m_i,$ $i=1,2,$ of
the two fermionic bound-state constituents are then interpreted as some
effective ones.\item The two bound-state constituents have equal masses $m,$
that is, $m_1=m_2=m.$ (The generalization to the case of unequal masses of
the bound-state constituents is, of course, straightforward \cite{resag94}.)
\end{itemize}Moreover, we illustrate the developed technique by investigating
mesons carrying the pion's quantum numbers, treated as bound states of a
quark--antiquark pair \cite{lucha91}.~Thus, we set up the instantaneous
Bethe--Salpeter equation for ${}^1{\rm S}_0$ bound states of massless
constituents with confining interaction kernels of time-component
Lorentz-vector~type.

\subsection{Pseudoscalar Bound States of Massless
Constituents}\label{Sec:pseudoscalar}We consider fermion--antifermion bound
states of total spin $J,$ with parity and charge-conjugation quantum numbers
$P$ and $C,$ expressed in terms of $J,$ given by $P=(-1)^{J+1}$ and
$C=(-1)^J,$ respectively. In usual spectroscopic notation, these states are
denoted by ${}^1J_J.$ The general expansion of the Salpeter amplitude $\chi$
in terms of a complete set of Dirac matrices performed in
Ref.~\cite{lagae92I} involves precisely eight independent components. For the
bound states under consideration, only two of these independent components
are relevant. We call the corresponding wave functions in momentum space
$\Psi_1$ and~$\Psi_2.$ In the notation of Laga\"e \cite{lagae92I}, these wave
functions are labelled $L_1$ and~$L_2,$ respectively.

For two particles of equal masses $m$ and internal momentum~$\bf k$
constituting a bound state with the quantum numbers specified above, the
Salpeter amplitude $\chi$ describing this bound state in its
center-of-momentum frame reads in momentum space$$\chi({\bf
k})=\left[\Psi_1({\bf k})\,\frac{m-\mbox{\boldmath{$\gamma$}}\cdot{\bf
k}}{E(k)}+\Psi_2({\bf k})\,\gamma^0\right]\gamma_5\ ,$$with the
abbreviation$$E(k)\equiv\sqrt{k^2+m^2}\ ,\quad k\equiv|{\bf k}|\ .$$The norm
$\|\chi\|$ of the Salpeter amplitude $\chi$ may be calculated from Eq.~(2.9)
of Ref.~\cite{lagae92I} (or from Eq.~(9) of Ref.~\cite{olsson95}). For the
above amplitude, the squared norm is given~by$$\|\chi\|^2=4\int\frac{{\rm
d}^3k}{(2\pi)^3}\,[\Psi_1^\ast({\bf k})\,\Psi_2({\bf k})+\Psi_2^\ast({\bf
k})\,\Psi_1({\bf k})]\ ,$$in accordance with Eq.~(4.13) of
Ref.~\cite{lagae92I}, and with Eq.~(18) of Ref.~\cite{olsson95}.

For simplicity, we focus our interest to quark--antiquark bound states of
spin $J=0;$ these are bound states with the spin-parity-charge conjugation
assignment $J^{PC}=0^{-+}.$ It is these pseudoscalar quark--antiquark bound
states which will be investigated here.

In view of the findings recalled in the Introduction, let us settle on the
investigation of the instantaneous Bethe--Salpeter equation for some
time-component Lorentz vector interaction between the bound-state
constituents, that is, on a Bethe--Salpeter kernel of the Dirac structure
$\gamma^0\otimes\gamma^0.$ For this case, the radial wave functions
$\Psi_1(k)$ and~$\Psi_2(k),$ obtained from $\Psi_1({\bf k})$ and~$\Psi_2({\bf
k})$ by factorizing off the spherical harmonics involving~the angular
variables, satisfy the following system of equations---which is equivalent to
the instantaneous Bethe--Salpeter equation and which may be deduced, e.g.,
from Eq.~(5.7) or Eq.~(5.9) of Ref.~\cite{lagae92I}, or from Eq.~(A1) of
Ref.~\cite{olsson95}:\begin{eqnarray}
&&2\,E(k)\,\Psi_2(k)+\int\limits_0^\infty\frac{{\rm d}k'\,k'^2}{(2\pi)^2}
\,V_0(k,k')\,\Psi_2(k')=M\,\Psi_1(k)\
,\nonumber\\[1ex]&&2\,E(k)\,\Psi_1(k)\label{Eq:IBSE}\\[1ex]&&
+\int\limits_0^\infty\frac{{\rm d}k'\,k'^2}{(2\pi)^2}
\left[\frac{m}{E(k)}\,V_0(k,k')\,\frac{m}{E(k')}
+\frac{k}{E(k)}\,V_1(k,k')\,\frac{k'}{E(k')}\right]\Psi_1(k')= M\,\Psi_2(k)\
,\nonumber\end{eqnarray}where, in terms of some interaction potential $V(r)$
in configuration space,\begin{equation}V_L(k,k')\equiv
8\pi\int\limits_0^\infty{\rm d}r\,r^2\,V(r)\,j_L(k\,r)\,j_L(k'\,r)\ ,\quad
L=0,1,2,\dots\ .\label{Eq:IBSE-intpot}\end{equation}Here, $j_n(z)$ ($n=0,\pm
1,\pm2,\dots$) are the spherical Bessel functions of the first kind
\cite{Abramowitz}.

For a vanishing mass of the two bound-state constituents (that is, for
$m=0$), this set of equations simplifies to\begin{eqnarray}
2\,k\,\Psi_2(k)+\frac{1}{(2\pi)^2}\int\limits_0^\infty{\rm d}k'\,k'^2\,
V_0(k,k')\,\Psi_2(k')&=&M\,\Psi_1(k)\ ,\nonumber\\[1ex]2\,k\,\Psi_1(k)
+\frac{1}{(2\pi)^2}\int\limits_0^\infty{\rm d}k'\,k'^2\,
V_1(k,k')\,\Psi_1(k')&=&M\,\Psi_2(k)\ .\label{Eq:IBSE-m=0}\end{eqnarray}The
above set of equations constitutes the starting point for the present
investigation.

\subsection{Analytical Solution by Series Expansion}\label{Sec:IBSE-sol}The
particular structure of the instantaneous Bethe--Salpeter equation
(\ref{Eq:IBSE}) allows us~to extract its solutions by the following simple
procedure. Express one of the components $\Psi_1,$ $\Psi_2$ of the Salpeter
amplitude $\chi$ from one of the coupled equations in the
system~(\ref{Eq:IBSE}) by the other component. Insert this expression into
the other equation of this system, in order to obtain an eigenvalue equation
for the other component with eigenvalue~$M^2$:
\begin{eqnarray*}M^2\,\Psi_2(k)&=&4\,k^2\,\Psi_2(k)\\[1ex]
&+&2\,k\int\limits_0^\infty\frac{{\rm d}k'\,k'^2}{(2\pi)^2}\,
V_0(k,k')\,\Psi_2(k')\\[1ex]&+&2\int\limits_0^\infty\frac{{\rm
d}k'\,k'^3}{(2\pi)^2}\, V_1(k,k')\,\Psi_2(k')\\[1ex]
&+&\int\limits_0^\infty\frac{{\rm d}k'\,k'^2}{(2\pi)^2}\,V_1(k,k')
\int\limits_0^\infty\frac{{\rm d}k''\,k''^2}{(2\pi)^2}\,V_0(k',k'')\,
\Psi_2(k'')\ .\end{eqnarray*}Clearly, all our results should be independent
of which component has been selected in the first place. For the present
analysis, we have chosen to express $\Psi_1$ in terms~of~$\Psi_2.$

An approximate solution to the above eigenvalue equation may be very
easily~found by expanding the ``Salpeter components'' over a suitable set of
basis functions \cite{lagae92I,resag94,olsson95}. To this end, we introduce
complete orthonormal systems $\{|\phi_i\rangle,i=0,1,2,\dots\}$ as bases for
the Hilbert space $L_2(R^+)$ of (with the weight function $w(x)=x^2$)
square-integrable functions $f(x)$ on the positive real line $R^+;$ these
basis vectors $|\phi_i\rangle$
satisfy~$\langle\phi_i|\phi_j\rangle=\delta_{ij}.$ The different sets of
basis vectors are distinguished by some parameter $\ell=0,1,2,\dots.$ The
representations of any basis vector $|\phi_i\rangle$ in configuration and
momentum space~are related to the corresponding bases for the Hilbert space
$L_2(R^3)$ of all square-integrable functions on the three-dimensional space
$R^3$ by factorizing off the spherical harmonics ${\cal Y}_{\ell m}(\Omega)$
for angular momentum $\ell=0,1,2,\dots$ and its projection
$m=-\ell,-\ell+1,\dots,+\ell$ which describe the dependence on the angular
variables summarized by the solid angle $\Omega.$ The parameter $\ell$ is
identified with the angular momentum of the $R^3$ basis functions. For a
given value of $\ell,$ the $R^+$ basis functions in configuration and
momentum space are called $\phi_i^{(\ell)}(r)$ and $\phi_i^{(\ell)}(p),$
respectively. They satisfy the orthonormalization conditions$$
\int\limits_0^\infty{\rm d}r\,r^2\,\phi_i^{\ast(\ell)}(r)\,\phi_j^{(\ell)}(r)=
\int\limits_0^\infty{\rm d}p\,p^2\,\phi_i^{\ast(\ell)}(p)\,\phi_j^{(\ell)}(p)=
\delta_{ij}\ ,\quad i,j=0,1,2,\dots\ .$$

In order to introduce an additional degree of freedom in the search for the
solutions of the instantaneous Bethe--Salpeter equation, we allow the basis
functions to depend on some positive real variational parameter $\mu$ (with
the dimension of mass). Our choice of radial basis functions, involving, in
their configuration-space representation~$\phi_i^{(\ell)}(r),$ the
generalized Laguerre polynomials \cite{Abramowitz}, is summarized in due
detail in Appendix~\ref{App:Laguerre}. A particular feature of this choice is
our phase convention: all configuration-space basis functions
$\phi_i^{(\ell)}(r)$ and, for $\ell$ even, the momentum-space basis functions
$\phi_i^{(\ell)}(p)$ are real. Starting with expansions over the radial basis
functions $\phi_i^{(0)}(p)$ corresponding to $\ell=0,$ the solution of the
instantaneous Bethe--Salpeter equation then simply amounts to the
diagonalization of the---{\it a priori}, infinite-dimensional---matrix (with
eigenvalues $M^2$)\begin{equation}{\cal M}_{ij}=A_{ij}+B_{ij}+C_{ij}+D_{ij}\
,\label{Eq:IBSE-matrix}\end{equation}with the abbreviations\begin{eqnarray}
A_{ij}&\equiv&4\int\limits_0^\infty{\rm d}k\,k^4\,\phi_i^{(0)}(k)\,
\phi_j^{(0)}(k)\ ,\label{Eq:IBSE-term-I}\\[1ex]
B_{ij}&\equiv&2\int\limits_0^\infty{\rm d}k\,k^3\,\phi_i^{(0)}(k)
\int\limits_0^\infty\frac{{\rm d}k'\,k'^2}{(2\pi)^2}\,
V_0(k,k')\,\phi_j^{(0)}(k')\ ,\label{Eq:IBSE-term-II}\\[1ex]
C_{ij}&\equiv&2\int\limits_0^\infty{\rm d}k\,k^2\,\phi_i^{(0)}(k)
\int\limits_0^\infty\frac{{\rm d}k'\,k'^3}{(2\pi)^2}\,V_1(k,k')\,
\phi_j^{(0)}(k')\ ,\label{Eq:IBSE-term-III}\\[1ex]
D_{ij}&\equiv&\int\limits_0^\infty{\rm d}k\,k^2\,\phi_i^{(0)}(k)
\int\limits_0^\infty\frac{{\rm d}k'\,k'^2}{(2\pi)^2}\,V_1(k,k')
\int\limits_0^\infty\frac{{\rm d}k''\,k''^2}{(2\pi)^2}\,V_0(k',k'')\,
\phi_j^{(0)}(k'')\ .\label{Eq:IBSE-term-IV}\end{eqnarray}In actual
calculations, an infinite-dimensional matrix, arising from a countably
infinite number of basis states $|\phi_i\rangle,$ is not manageable. One has
to be content with a truncation of the matrix ${\cal M}_{ij}$ to a, say,
$d\times d$ matrix. Moreover, in the expansions encountered~in the
intermediate steps only the first, say, $N+1$ basis vectors can be taken into
account.

\newpage\noindent The (straightforward) evaluation of the matrix ${\cal
M}_{ij}$ is briefly sketched in Appendix~\ref{App:conversion}:
\begin{eqnarray*}{\cal M}_{ij}&=&4\,I^{(2)}_{ij}(\mu)
+2\,\sum_{r=0}^N\,b_{ri}(\mu)\,V^{(0)}_{rj}(\mu)
+2\,\sum_{r=0}^N\,\sum_{s=0}^N\,c^\ast_{ri}\,d_{sj}(\mu)\,V^{(1)}_{rs}(\mu)
\nonumber\\[1ex]&+&\sum_{r=0}^N\,\sum_{s=0}^N\,\sum_{t=0}^N\,c^\ast_{ri}\,
c_{st}\,V^{(1)}_{sr}(\mu)\,V^{(0)}_{tj}(\mu)\ .\end{eqnarray*}Let us present
the various quantities entering in this result in their order of appearance.
The (real and symmetric) matrix $I^{(2)}_{ij}(\mu)$ may be easily read off
from the result~(\ref{Eq:nth-moment-result})~for the integral
$I^{(n)}_{ij}(\mu)$ defined by Eq.~(\ref{Eq:nth-moment}) for the case
$n=2$:\begin{eqnarray*}
I^{(2)}_{ij}(\mu)&=&\frac{4\,\mu^2}{\pi\,\sqrt{(i+1)\,(i+2)\,(j+1)\,(j+2)}}
\\[1ex]&\times&\sum_{r=0}^i\,\sum_{s=0}^j\,(-2)^{r+s}
\left(\begin{array}{c}i+2\\i-r\end{array}\right)
\left(\begin{array}{c}j+2\\j-s\end{array}\right)(r+1)\,(s+1)\\[1ex]&\times&
\left[\sum_{k=0}^{|r-s|}\left(\begin{array}{c}|r-s|\\k\end{array}\right)
\frac{\Gamma(\frac{1}{2}\,(k+3))\,\Gamma(\frac{1}{2}\,(r+s+1+|r-s|-k))}
{\Gamma(\frac{1}{2}\,(r+s+4+|r-s|))}
\cos\left(\frac{k\,\pi}{2}\right)\right.\\[1ex]
&-&\left.\sum_{k=0}^{r+s+4}\left(\begin{array}{c}r+s+4\\k\end{array}\right)
\frac{\Gamma(\frac{1}{2}\,(k+3))\,\Gamma(\frac{1}{2}\,(2\,r+2\,s+5-k))}
{\Gamma(r+s+4)}\cos\left(\frac{k\,\pi}{2}\right)\right].\end{eqnarray*}Note
the scaling behaviour of this expression with respect to the variational
parameter $\mu$:$$I^{(2)}_{ij}(\mu)=\mu^2\,I^{(2)}_{ij}(1)\ .$$Explicitly,
the matrix $I^{(2)}(\mu)\equiv\left(I^{(2)}_{ij}(\mu)\right)$ reads
$$I^{(2)}(\mu)\equiv\left(I^{(2)}_{ij}(\mu)\right)=
\mu^2\left(\begin{array}{ccc}1&\displaystyle\frac{2}{\sqrt{3}}&\cdots\\[3ex]
\displaystyle\frac{2}{\sqrt{3}}&\displaystyle\frac{7}{3}&\cdots\\[3ex]
\vdots&\vdots&\ddots\end{array}\right).$$According to Eq.~(\ref{Eq:b=I1}),
the real and symmetric matrix $b_{ij}(\mu)$ is identical to the integral
$I^{(n)}_{ij}(\mu)$ defined by Eq.~(\ref{Eq:nth-moment}) for the case $n=1;$
consequently, it may be read off from~the result~(\ref{Eq:nth-moment-result})
for $n=1$:\begin{eqnarray*}b_{ij}(\mu)&=&
\frac{4\,\mu}{\pi\,\sqrt{(i+1)\,(i+2)\,(j+1)\,(j+2)}}\\[1ex]
&\times&\sum_{r=0}^i\,\sum_{s=0}^j\,(-2)^{r+s}
\left(\begin{array}{c}i+2\\i-r\end{array}\right)
\left(\begin{array}{c}j+2\\j-s\end{array}\right)(r+1)\,(s+1)\\[1ex]&\times&
\left[\sum_{k=0}^{|r-s|}\left(\begin{array}{c}|r-s|\\k\end{array}\right)
\frac{\Gamma(\frac{1}{2}\,(k+2))\,\Gamma(\frac{1}{2}\,(r+s+2+|r-s|-k))}
{\Gamma(\frac{1}{2}\,(r+s+4+|r-s|))}
\cos\left(\frac{k\,\pi}{2}\right)\right.\\[1ex]
&-&\left.\sum_{k=0}^{r+s+4}\left(\begin{array}{c}r+s+4\\k\end{array}\right)
\frac{\Gamma(\frac{1}{2}\,(k+2))\,\Gamma(\frac{1}{2}\,(2\,r+2\,s+6-k))}
{\Gamma(r+s+4)}\cos\left(\frac{k\,\pi}{2}\right)\right].\end{eqnarray*}Note
the scaling behaviour of this expression with respect to the variational
parameter $\mu$:$$b_{ij}(\mu)=\mu\,b_{ij}(1)\ .$$Explicitly, the matrix
$b(\mu)\equiv(b_{ij}(\mu))$
reads$$b(\mu)\equiv(b_{ij}(\mu))=\frac{8\,\mu}{3\pi}\left(
\begin{array}{ccc}1&\displaystyle\frac{1}{\sqrt{3}}&\cdots\\[3ex]
\displaystyle\frac{1}{\sqrt{3}}&\displaystyle\frac{7}{5}&\cdots\\[3ex]
\vdots&\vdots&\ddots\end{array}\right).$$According to Eq.~(\ref{Eq:c=J0}),
the (purely imaginary) matrix $c_{ij}$ is nothing else but the integral
$J^{(n)}_{ij}(\mu)$ defined by Eq.~(\ref{Eq:nth-overlap}) specified to the
case $n=0;$ consequently, it may be read~off from the
result~(\ref{Eq:nth-overlap-result}) for $n=0$ (here, because of the
singularity of $\Gamma(0),$ some care has~to be taken when performing the
limit $n\to 0$ in the algebraic expression~(\ref{Eq:nth-overlap-result}) for
$J^{(n)}_{ij}(\mu)$):\begin{eqnarray*}c_{ij}&=&{\rm
i}\,\frac{8}{\pi\,\sqrt{(i+1)\,(i+2)\,(i+3)\,(i+4)\,(j+1)\,(j+2)}}\\[1ex]
&\times&\sum_{r=0}^i\,\sum_{s=0}^j\,(-2)^{r+s}\,(r+1)\,(r+2)\,(r+3)\,(s+1)
\left(\begin{array}{c}i+4\\i-r\end{array}\right)
\left(\begin{array}{c}j+2\\j-s\end{array}\right)\\[1ex]
&\times&\left\{\frac{1}{r+2}\left[-\Psi(\mbox{$\frac{1}{2}$}\,(4+r+s+|r-s|))+
\Psi(4+r+s)\frac{}{}\right.\right.\\[1ex]
&+&\sum_{k=2}^{|r-s|}\left(\begin{array}{c}|r-s|\\k\end{array}\right)
\frac{\Gamma(\frac{1}{2}\,k)\,\Gamma(\frac{1}{2}\,(4+r+s+|r-s|-k))}
{\Gamma(\frac{1}{2}\,(4+r+s+|r-s|))}\cos\left(\frac{k\,\pi}{2}\right)\\[1ex]
&-&\left.\sum_{k=2}^{4+r+s}\,\left(\begin{array}{c}4+r+s\\k\end{array}\right)
\frac{\Gamma(\frac{1}{2}\,k)\,\Gamma(\frac{1}{2}\,(8+2\,r+2\,s-k))}
{\Gamma(4+r+s)}\cos\left(\frac{k\,\pi}{2}\right)\right]\\[1ex]
&-&\frac{1}{r+3}\left[
-\Psi(\mbox{$\frac{1}{2}$}\,(5+r+s+|1+r-s|))+\Psi(5+r+s)\frac{}{}\right.\\[1ex]
&+&\sum_{k=2}^{|1+r-s|}\left(\begin{array}{c}|1+r-s|\\k\end{array}\right)
\frac{\Gamma(\frac{1}{2}\,k)\,\Gamma(\frac{1}{2}\,(5+r+s+|1+r-s|-k))}
{\Gamma(\frac{1}{2}\,(5+r+s+|1+r-s|))}\cos\left(\frac{k\,\pi}{2}\right)\\[1ex]
&-&\left.\left.\sum_{k=2}^{5+r+s}
\left(\begin{array}{c}5+r+s\\k\end{array}\right)
\frac{\Gamma(\frac{1}{2}\,k)\,\Gamma(\frac{1}{2}\,(10+2\,r+2\,s-k))}
{\Gamma(5+r+s)}\cos\left(\frac{k\,\pi}{2}\right)\right]\right\},\end{eqnarray*}
where $\Psi(z)$ denotes the logarithmic derivative of the gamma function, the
``digamma function.'' Note that the quantities $c_{ij}$ are independent of
the variational parameter~$\mu.$ Explicitly, the matrix $c\equiv(c_{ij})$
reads$$c\equiv(c_{ij})={\rm i}\,\frac{16}{3\pi}\left(
\begin{array}{ccc}\displaystyle\frac{1}{\sqrt{3}}
&\displaystyle\frac{1}{15}&\cdots\\[3ex]-\displaystyle\frac{1}{5\,\sqrt{15}}&
\displaystyle\frac{19}{15\,\sqrt{5}}&\cdots\\[3ex]
\vdots&\vdots&\ddots\end{array}\right).$$Similarly, according to
Eq.~(\ref{Eq:d=J1}), the (purely imaginary) matrix $d_{ij}(\mu)$ is
identical~to~the integral $J^{(n)}_{ij}(\mu)$ defined by
Eq.~(\ref{Eq:nth-overlap}) specified to the case $n=1;$ consequently, it
may~be read off from the result~(\ref{Eq:nth-overlap-result}) for
$n=1$:\begin{eqnarray*}&&d_{ij}(\mu)\\[1ex]&&={\rm
i}\,\frac{8\,\mu}{\pi\,\sqrt{(i+1)\,(i+2)\,(i+3)\,(i+4)\,(j+1)\,(j+2)}}\\[1ex]
&&\times\sum_{r=0}^i\,\sum_{s=0}^j\,(-2)^{r+s}\,(r+1)\,(r+2)\,(r+3)\,(s+1)
\left(\begin{array}{c}i+4\\i-r\end{array}\right)
\left(\begin{array}{c}j+2\\j-s\end{array}\right)\\[1ex]
&&\times\left\{\frac{1}{r+2}\left[\sum_{k=0}^{|r-s|}
\left(\begin{array}{c}|r-s|\\k\end{array}\right)
\frac{\Gamma(\frac{1}{2}\,(k+1))\,\Gamma(\frac{1}{2}\,(3+r+s+|r-s|-k))}
{\Gamma(\frac{1}{2}\,(4+r+s+|r-s|))}
\cos\left(\frac{k\,\pi}{2}\right)\right.\right.\\[1ex]
&&-\left.\sum_{k=0}^{4+r+s}\,\left(\begin{array}{c}4+r+s\\k\end{array}\right)
\frac{\Gamma(\frac{1}{2}\,(k+1))\,\Gamma(\frac{1}{2}\,(7+2\,r+2\,s-k))}
{\Gamma(4+r+s)}\cos\left(\frac{k\,\pi}{2}\right)\right]\\[1ex]
&&-\frac{1}{r+3}\left[\sum_{k=0}^{|1+r-s|}
\left(\begin{array}{c}|1+r-s|\\k\end{array}\right)\right.\\[1ex]&&\times
\frac{\Gamma(\frac{1}{2}\,(k+1))\,\Gamma(\frac{1}{2}\,(4+r+s+|1+r-s|-k))}
{\Gamma(\frac{1}{2}\,(5+r+s+|1+r-s|))}\cos\left(\frac{k\,\pi}{2}\right)
\\[1ex]&&\left.\left.-\sum_{k=0}^{5+r+s}
\left(\begin{array}{c}5+r+s\\k\end{array}\right)
\frac{\Gamma(\frac{1}{2}\,(k+1))\,\Gamma(\frac{1}{2}\,(9+2\,r+2\,s-k))}
{\Gamma(5+r+s)}\cos\left(\frac{k\,\pi}{2}\right)\right]\right\}.\end{eqnarray*}
Note the scaling behaviour of this expression with respect to the variational
parameter $\mu$:$$d_{ij}(\mu)=\mu\,d_{ij}(1)\ .$$Explicitly, the matrix
$d(\mu)\equiv(d_{ij}(\mu))$ reads$$d(\mu)\equiv(d_{ij}(\mu))={\rm
i}\,\frac{\mu}{2}\left(\begin{array}{ccc}\sqrt{3}&1&\cdots\\[3ex]
\sqrt{\displaystyle\frac{3}{5}}&\sqrt{5}&\cdots\\[3ex]\vdots&\vdots&\ddots
\end{array}\right).$$Finally, the real and symmetric matrix
$V^{(\ell)}_{ij}(\mu)$ of expectation values of the interaction potential
$V(r)$ with respect to the basis states $|\phi_i\rangle$ (characterized by a
particular value of $\ell$) is defined by\begin{equation}V^{(\ell)}_{ij}(\mu)
\equiv\langle\phi_i|V(r)|\phi_j\rangle=\int\limits_0^\infty{\rm
d}r\,r^2\,V(r)\,\phi_i^{(\ell)}(r)\,\phi_j^{(\ell)}(r)\
.\label{Eq:expval-intpot}\end{equation}For power-law potentials, that is, for
interaction potentials $V(r)$ of the power-law form
$$V(r)=\sum_na_n\,r^{b_n}\ ,$$with (arbitrary) real constants $a_n$ and
$b_n,$ these matrix elements $\langle\phi_i|V(r)|\phi_j\rangle$ are easily
worked out algebraically \cite{Lucha97,Lucha98O,Lucha98D}. The algebraic
expression for the general case can be found in Sec.~4 of
Ref.~\cite{Lucha97}, Sec.~3.10 of Ref.~\cite{Lucha98O}, or Sec.~2.8.1 of
Ref.~\cite{Lucha98D}.

The simplest model for a confining interaction between all bound-state
constituents is provided by a linear potential\begin{equation}V(r)=
\lambda\,r\ ,\quad\lambda>0\ ,\label{Eq:linpot}\end{equation}
whence\begin{equation}V_L(k,k')=8\pi\,\lambda\int\limits_0^\infty{\rm
d}r\,r^3\,j_L(k\,r)\,j_L(k'\,r)\ ,\quad L=0,1,2,\dots\
.\label{Eq:linpot-Bessel}\end{equation}In this case, the above-mentioned
general expression for the expectation values $V^{(\ell)}_{ij}(\mu)$ of the
interaction potential $V(r)$, taken with respect to our basis states
$|\phi_i\rangle,$ simplifies to\begin{eqnarray*}
V^{(\ell)}_{ij}(\mu,\lambda)&=&\sqrt{\frac{i!\,j!}{\Gamma(2\,\ell+i+3)\,
\Gamma(2\,\ell+j+3)}}\,\frac{\lambda}{2\,\mu}\,\sum_{r=0}^i\,\sum_{s=0}^j\,
\frac{(-1)^{r+s}}{r!\,s!}\\[1ex]&\times&
\left(\begin{array}{c}i+2\,\ell+2\\i-r\end{array}\right)
\left(\begin{array}{c}j+2\,\ell+2\\j-s\end{array}\right)\Gamma(2\,\ell+r+s+4)\
.\end{eqnarray*}Note the scaling behaviour of the potential matrix
$V^{(\ell)}_{ij}(\mu,\lambda),$ $\ell=0,1,\dots,$ with respect to the
variational parameter $\mu$ and the coupling strength $\lambda$ of the linear
potential (\ref{Eq:linpot}):$$V^{(\ell)}_{ij}(\mu,\lambda)=
\frac{\lambda}{\mu}\,V^{(\ell)}_{ij}(1,1)\ .$$The explicit expression of this
potential matrix $V^{(\ell)}(\mu,\lambda)$ reads, for instance, for
$\ell=0,$$$V^{(0)}(\mu,\lambda)\equiv\left(V^{(0)}_{ij}(\mu,\lambda)\right)=
\frac{\lambda}{2\,\mu}\left(\begin{array}{ccc}3&-\sqrt{3}&\cdots\\[1ex]
-\sqrt{3}&5&\cdots\\[1ex]\vdots&\vdots&\ddots\end{array}\right)$$and, for
$\ell=1,$$$V^{(1)}(\mu,\lambda)\equiv\left(V^{(1)}_{ij}(\mu,\lambda)\right)=
\frac{\lambda}{2\,\mu}\left(\begin{array}{ccc}5&-\sqrt{5}&\cdots\\[1ex]
-\sqrt{5}&7&\cdots\\[1ex]\vdots&\vdots&\ddots\end{array}\right).$$

Taking into account, for the special case of a linear potential, the
above-mentioned scaling behaviour of all the expressions entering in our main
result for the matrix ${\cal M}_{ij},$ we finally arrive
at\begin{eqnarray}{\cal M}_{ij}&=&4\,\mu^2\,I^{(2)}_{ij}(1)
+2\,\lambda\,\sum_{r=0}^N\,b_{ri}(1)\,V^{(0)}_{rj}(1,1)
+2\,\lambda\,\sum_{r=0}^N\,\sum_{s=0}^N\,c^\ast_{ri}\,d_{sj}(1)\,
V^{(1)}_{rs}(1,1)\nonumber\\[1ex]
&+&\frac{\lambda^2}{\mu^2}\,\sum_{r=0}^N\,\sum_{s=0}^N\,\sum_{t=0}^N\,
c^\ast_{ri}\,c_{st}\,V^{(1)}_{sr}(1,1)\,V^{(0)}_{tj}(1,1)\
.\label{Eq:IBSE-matrix-scaled}\end{eqnarray}Clearly, this structure of the
matrix ${\cal M}_{ij},$ with the squared masses of the bound states as its
eigenvalues, has to emerge already on dimensional grounds: the mass dimension
of the variational parameter $\mu$ is 1 while the mass dimension of the
slope~$\lambda$ of the~linear potential~(\ref{Eq:linpot}) is 2; the potential
$V(r)$ enters into the expression (\ref{Eq:IBSE-matrix}) for the
matrix~${\cal M}_{ij}$ in, at most, second order.

\section{Results}The dependence of the matrix ${\cal M}_{ij}$ on the
variational parameter $\mu,$ as apparent from
Eq.~(\ref{Eq:IBSE-matrix-scaled}), is
\begin{equation}{\cal M}_{ij}=\mu^2\,E_{ij}+F_{ij}+\frac{1}{\mu^2}\,G_{ij}\
,\label{Eq:IBSE-matrix-mu}\end{equation}with\begin{eqnarray*}E_{ij}&\equiv&
4\,I^{(2)}_{ij}(1)\ ,\\[1ex]
F_{ij}&\equiv&2\,\lambda\,\sum_{r=0}^N\,b_{ri}(1)\,V^{(0)}_{rj}(1,1)
+2\,\lambda\,\sum_{r=0}^N\,\sum_{s=0}^N\,c^\ast_{ri}\,d_{sj}(1)\,
V^{(1)}_{rs}(1,1)\ ,\\[1ex]
G_{ij}&\equiv&\lambda^2\,\sum_{r=0}^N\,\sum_{s=0}^N\,\sum_{t=0}^N\,
c^\ast_{ri}\,c_{st}\,V^{(1)}_{sr}(1,1)\,V^{(0)}_{tj}(1,1)\ .\end{eqnarray*}

Any solution method relying on truncated expansions over some complete
set(s)~of basis vectors only makes sense if the approximate result obtained
in this way converges to the exact solution of the problem under study for
increasing numbers of basis vectors taken into account in the expansions.
Assuming this convergence to hold, a reasonable estimate of the uncertainties
induced by the truncation of the expansion series to finite numbers of basis
vectors can be given by the inspection of the lowest-dimensional case. For
$i=j=0,$ we find, for the first elements of the matrices $I^{(2)}(1),$
$b(1),$ $c,$ and $d(1),$$$I^{(2)}_{00}(1)=1\ ,\quad b_{00}(1)=\frac{8}{3\pi}\
,\quad c_{00}={\rm i}\,\frac{16}{3\,\sqrt{3}\,\pi}\ ,\quad d_{00}(1)={\rm
i}\,\frac{\sqrt{3}}{2}\ ,$$and, for the expectation values $V^{(\ell)}(1,1),$
$\ell=0,1,$ of the linear potential,$$\quad V^{(0)}_{00}(1,1)=\frac{3}{2}\
,\quad V^{(1)}_{00}(1,1)=\frac{5}{2}\ .$$This implies, for $N=0,$ for the
quantities $E\equiv E_{00},$ $F\equiv F_{00},$ and $G\equiv G_{00},$$$E=4\
,\quad F=\frac{64}{3\pi}\,\lambda\ ,\quad G=\frac{320}{9\,\pi^2}\,\lambda^2\
.$$Minimizing$${\cal M}_{00}=\mu^2\,E+F+\frac{1}{\mu^2}\,G$$with respect to
the variational parameter $\mu,$ yields$${\cal M}_{00}=2\,\sqrt{E\,G}+F\ .$$
The above values of $E,$ $F,$ and $G$ then yield, for the square of the
bound-state mass~$M$,\begin{equation}M^2=
\frac{32}{3\pi}\left(2+\sqrt{5}\right)\lambda\ .\label{Eq:M-1d}\end{equation}

For a linear interaction potential like the one used in this investigation,
it would~be natural to present the numerical results for the masses $M$ of
the bound states in terms of the dimensionless quantities $M/\sqrt{\lambda},$
as has been done, e.g., in Ref.~\cite{lagae92II}. However,~in order to
facilitate the comparison of our findings with the ones quoted in
Ref.~\cite{olsson95},~we present our results for the particular, more or less
reasonable \cite{lucha91} value of the slope~$\lambda$ of the linear
interaction potential (\ref{Eq:linpot}) adopted in Ref.~\cite{olsson95},
viz., for $\lambda=0.2\;\mbox{GeV}^2.$~For this slope we obtain, for~the
``one-dimensional'' result (\ref{Eq:M-1d}) for the bound-state mass~$M$,
$M=1.69604\;\mbox{GeV}.$

\newpage In general, the approximate values of the bound-state masses $M$,
calculated within the present approach as the square roots of the eigenvalues
of the matrix ${\cal M}_{ij},$ depend on the variational parameter $\mu.$ In
our almost entirely algebraic method of solving the instantaneous
Bethe--Salpeter equation, this dependence on $\mu$ has been made manifest in
form of Eq.~(\ref{Eq:IBSE-matrix-mu}). However, any solution technique
relying on some series expansions can be regarded as reasonable only if it
exhibits stability with respect to the increase~in the number of basis
states, in the sense that increasing the number of basis states---in our
case, increasing the size $d$ of the matrix ${\cal M}_{ij}$ or the number $N$
of basis states~taken into account in the series expansions performed in the
intermediate steps---reduces the dependence of the results, the bound-state
masses $M,$ on the variational parameter $\mu.$

\begin{figure}[h]
\begin{center}\psfig{figure=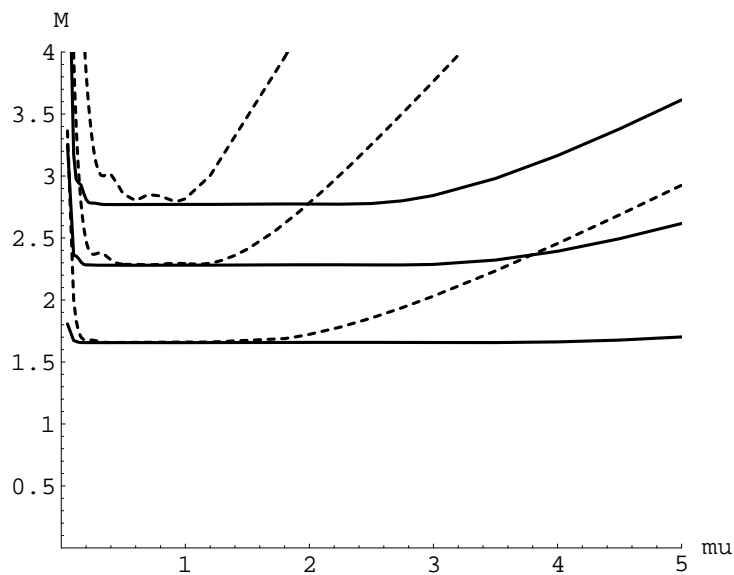,scale=0.955}\caption{\small
Dependence of the masses $M$ (in units of GeV) of the three lowest-lying
(positive-norm) $J^{PC}=0^{-+}$ bound states of two massless constituents
obtained by the instantaneous Bethe--Salpeter equation, with a time-component
Lorentz-vector confining interaction kernel involving a linear potential
$V(r)=\lambda\,r$ with slope $\lambda=0.2\;\mbox{GeV}^2$, on the variational
parameter $\mu$ (in units of GeV) for $N=49$ and matrix sizes $d=5$ (dashed
lines) and $d=15$ (full lines).}\label{Fig:masses}\end{center}\end{figure}

Figure~\ref{Fig:masses} shows (in perfect agreement with Fig.~1 of
Ref.~\cite{olsson95}) the dependence~of~the masses $M$ of the three
lowest-lying bound states, i.e., the first three radial excitations,
extracted from Eq.~(\ref{Eq:IBSE-matrix-mu}) for a linear interaction
potential (\ref{Eq:linpot}) with slope $\lambda=0.2\;\mbox{GeV}^2,$ on the
variational parameter $\mu.$ When increasing the number of basis states (by
varying the matrix size $d$ from $d=5$ to $d=15$) the formation of regions
where the eigenvalues $M$ become independent of $\mu$ should be observed;
this feature is indeed found in Fig.~\ref{Fig:masses}.

For the ground state, the region of $\mu$-independence corresponding to the
matrix~size $d=5$ is centered somewhere around $\mu=1\;\mbox{GeV}.$ For this
value of $\mu,$ the ground-state mass $M,$ calculated for a matrix size
$d=15$ and $N=49$ in the intermediate-step~series expansions, is
$M=1.656\;\mbox{GeV},$ which reveals that the rough
approximation~(\ref{Eq:M-1d}) of~the above one-dimensional analysis is not
too bad (at least for the situation studied here).

In order to pave the way for a future quantitative discussion of the accuracy
of~our method as well as for a numerical comparison with the findings of
different approaches, we list in Table~\ref{Tab:BS-masses-mudep} the masses
$M$ of the three lowest-lying bound states for several values of the
variational parameter $\mu$ in the neighborhood of the above ``region of
stability.''\begin{table}[ht]\caption{Eigenvalues $M,$ in units of GeV, of
the instantaneous Bethe--Salpeter equation for two massless (i.e., $m=0$)
spin-$\frac{1}{2}$ fermions experiencing an interaction described~by a linear
potential with slope $\lambda=0.2\;\mbox{GeV}^2$ and forming bound states of
radial quantum number $n_{\rm r}=0,1,2$ and spin-parity-charge conjugation
assignment $J^{PC}=0^{-+}$ (called $1^1{\rm S}_0,$ $2^1{\rm S}_0,$ and
$3^1{\rm S}_0$ in the usual spectroscopic notation) as functions of the
variational parameter $\mu,$ for matrix size $d=15$ and $N=49$ in the
intermediate series expansions.}\label{Tab:BS-masses-mudep}
\begin{center}\begin{tabular}{llll}\hline\hline&&\\[-1.5ex]
\multicolumn{1}{c}{$\mu$ [GeV]}&\multicolumn{1}{c}{$1^1{\rm S}_0$}&
\multicolumn{1}{c}{$2^1{\rm S}_0$}&\multicolumn{1}{c}{$3^1{\rm S}_0$}\\[1ex]
\hline\\[-1.5ex]
0.05&1.807&3.253&5.392\\0.1&1.673&2.360&3.173\\0.2&1.656&2.283&2.812\\
0.5&1.656&2.280&2.770\\1&1.656&2.281&2.771\\1.2&1.657&2.281&2.771\\
2&1.658&2.284&2.774\\3&1.658&2.287&2.843\\4&1.662&2.393&3.165\\
5&1.702&2.617&3.614\\[1ex]\hline\hline\end{tabular}\end{center}\end{table}

In our procedure, the series expansion of the radial Salpeter component
$\Psi_2(k)$ with respect to the (primary) system
$\{\phi_i^{(0)}(k),i=0,1,2,\dots\}$ of radial basis functions~reads
\begin{equation}\Psi_2(k)=\sum_{i=0}^{d-1}\,e_i\,\phi_i^{(0)}(k)\
.\label{Eq:Psi_2(k)}\end{equation}The expansion coefficients $e_i$ in this
series are obtained, in the course of diagonalization of the matrix ${\cal
M}_{ij}$ introduced in Eq.~(\ref{Eq:IBSE-matrix}), as the eigenvectors of
${\cal M}_{ij}.$ From the first of the two relations constituting the
instantaneous Bethe--Salpeter equation for massless constituents,
Eq.~(\ref{Eq:IBSE-m=0}), the radial Salpeter component $\Psi_1(k)$ is then
obtained in
the~form\begin{equation}\Psi_1(k)=\sum_{i=0}^{d-1}\,e_i\,\psi_i(k)\
,\label{Eq:Psi_1(k)}\end{equation}where---in order to match the structure of
the first of Eqs.~(\ref{Eq:IBSE-m=0})---we took the liberty~to {\em
define}$$\psi_i(k):=\frac{1}{M}\left[2\,k\,\phi_i^{(0)}(k)+\sum_{j=0}^N\,
V^{(0)}_{ji}(\mu)\,\phi_j^{(0)}(k)\right],\quad i=0,1,2,\dots\ .$$The
corresponding representations of the above Salpeter components $\Psi_1(k)$
and $\Psi_2(k)$ in configuration space, denoted by $\Psi_1(r)$ and
$\Psi_2(r),$ are calculated by a Fourier--Bessel transformation (for
$J=0$):\begin{eqnarray}
\Psi_1(r)&=&\sum_{i=0}^{d-1}\,e_i\,\psi_i(r)\ ,\label{Eq:Psi_1(r)}\\[1ex]
\Psi_2(r)&=&\sum_{i=0}^{d-1}\,e_i\,\phi_i^{(0)}(r)\ ,\label{Eq:Psi_2(r)}
\end{eqnarray}with$$\psi_i(r)=\frac{1}{M}\left[
2\,\sum_{j=0}^N\,b_{ji}(\mu)\,\phi_j^{(0)}(r)+V(r)\,\phi_i^{(0)}(r)\right],
\quad i=0,1,2,\dots\ .$$ Clearly, the relevant objects are the {\em matrix
elements\/} calculated from these amplitudes.

\begin{figure}[ht]\begin{center}\psfig{figure=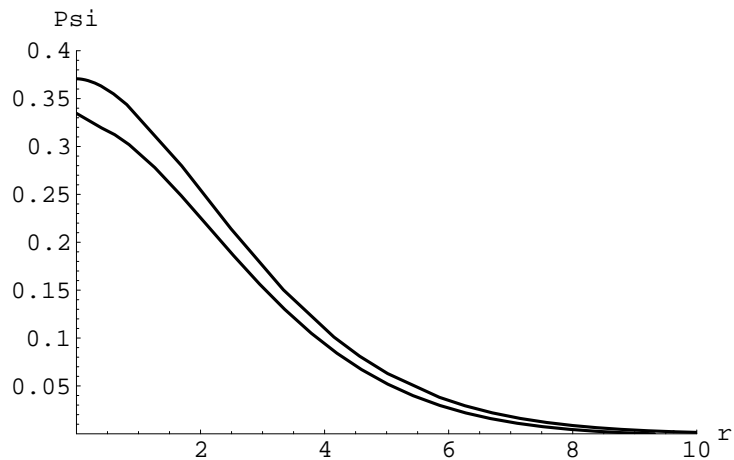,
scale=0.955}\caption{\small Radial ``Salpeter components'' $\Psi_1(r)$ and
$\Psi_2(r)$ in configuration space (in units of $\mbox{GeV}^{3/2}$) for the
lowest-lying (positive-norm) $J^{PC}=0^{-+}$ bound state of two massless
spin-$\frac{1}{2}$ constituents obtained by the instantaneous Bethe--Salpeter
equation, with a time-component Lorentz-vector confining interaction kernel
involving a linear potential $V(r)=\lambda\,r$ with slope
$\lambda=0.2\;\mbox{GeV}^2$, as a function of the relative distance $r$ (in
units of $\mbox{GeV}^{-1}$) of the bound-state constituents, for a matrix
size $d=50$ and $N=49$ in the intermediate-step series expansions; the lower
and upper curves represent the Salpeter components $\Psi_1(r)$ and
$\Psi_2(r),$ respectively.}\label{Fig:components}\end{center}\end{figure}

Figure~\ref{Fig:components} shows (similarly to Fig.~\ref{Fig:masses} in
reasonable agreement with Fig.~4 of Ref.~\cite{olsson95}, as judged by the
eye) the behaviour of the radial Salpeter components $\Psi_1(r)$ and
$\Psi_2(r)$ in configuration space for the ground state of the physical
system under consideration. According to Sec.~\ref{Sec:pseudoscalar}, the
normalization of the radial Salpeter components $\Psi_1(k)$ and $\Psi_2(k)$
in momentum space is determined by the norm $\|\chi\|$ of the Salpeter
amplitude~$\chi$:$$\|\chi\|^2=\frac{4}{(2\pi)^3}\int\limits_0^\infty{\rm
d}k\,k^2\,[\Psi_1^\ast(k)\,\Psi_2(k)+\Psi_2^\ast(k)\,\Psi_1(k)]\ .$$In the
present approach (summarized, as far as the Salpeter amplitude $\chi$ is
concerned, in Eqs.~(\ref{Eq:Psi_1(k)}) and (\ref{Eq:Psi_2(k)}) or
Eqs.~(\ref{Eq:Psi_1(r)}) and (\ref{Eq:Psi_2(r)}), respectively) the
normalization of the~radial component $\Psi_2$ is identical in configuration
and momentum space:$$\int\limits_0^\infty{\rm d}r\,r^2\,|\Psi_2(r)|^2
=\int\limits_0^\infty{\rm d}k\,k^2\,|\Psi_2(k)|^2=\sum_{i=0}^{d-1}\,|e_i|^2\
.$$For the plot presented in Fig.~\ref{Fig:components}, we have chosen this
normalization to be equal to~unity. The normalization of the radial component
$\Psi_1$ then follows from Eq.~(\ref{Eq:Psi_1(r)}) or
Eq.~(\ref{Eq:Psi_1(k)}). Once the desired value of the norm $\|\chi\|$ of the
Salpeter amplitude $\chi$ has been fixed,~it~is an easy task to determine the
common normalization factor of $\Psi_1$ and $\Psi_2$ such that the Salpeter
amplitude involving these wave functions satisfies its normalization
condition.

\section{Summary, Conclusions, and Outlook}The present investigation has been
devoted to the explicit construction of some matrix representation of the
Bethe--Salpeter equation in the instantaneous approximation for the involved
interaction kernel. As usual, this conversion to a matrix equation has~been
achieved by systematic expansions of the solutions and certain intermediate
quantities in terms of suitably chosen sets of basis states in the Hilbert
space under~consideration. A very important feature of our particular choice
of basis states is that these states~may be represented equally well in
configuration and in momentum space. As a consequence of this, matrix
elements of operators may be evaluated in the (with respect to the~given
operator) more convenient representation space: the matrix elements of
the~relativistic kinetic-energy operator (which, because of the troublesome
square root, is nonlocal~in configuration space) can be evaluated in momentum
space (where it is represented~by~a simple multiplication operator which, of
course, will be reminiscent of the square~root); on~the other hand, the
matrix elements of the interaction-potential operator (which~is, in general,
nonlocal in momentum space) can be evaluated in configuration space (that is,
the environment where the interaction potentials are more frequently
formulated). For the wide class of power-law potentials, the algebraic
expression of the latter matrix elements can be easily deduced from the
general result given in any of Refs.~\cite{Lucha97,Lucha98O,Lucha98D}. Merely
for the sake of illustration of our technique, we adopted the linear
potential~as some kind of toy model for quark confinement.

Our main result is the formulation of the instantaneous Bethe--Salpeter
equation~as a matrix eigenvalue problem {\em with explicitly known matrices}.
We notice several highly welcome features of this approach, which render it a
very efficient technique of solution:\begin{enumerate}\item Because of the
scaling behaviour of the involved quantities pointed out in
Sec.~\ref{Sec:IBSE-sol} at every instance, all dependence on both the
variational parameter $\mu$ and on~the coupling constants entering into the
interaction potential $V(r)$ (like, in the case of a linear potential, the
slope $\lambda$) may be factored out from the matrix elements. As a
consequence, only matrices evaluated for unit values of the parameters are
needed in our explicit formulation of the eigenvalue problem. For a
given~number of basis states taken into account, these matrices have to be
calculated only~once. This set of matrices includes the ``moment'' integrals
$I^{(2)}_{ij}(1),$ all the (non-square) matrices of expansion coefficients
$b_{ij}(1),$ $c_{ij},$ and $d_{ij}(1),$ as well as the expectation values
$V^{(\ell)}_{ij}(1,1),$ $\ell=0,1,$ of the linear interaction potential (in
our special case).\item The only numerical operation required in the course
of our method of solution is the diagonalization of the matrix representing
the instantaneous Bethe--Salpeter equation. For the size of this matrix
representation small enough, more precisely, for $d\le 4,$ even this
diagonalization may be performed analytically. Because of~an adequate choice
of basis, already the case $d=1$ yields a remarkably good
result.\end{enumerate}Thus, our analytic approach offers a viable way of
solving the Bethe--Salpeter equation.

Finally, we feel obliged to address the questions of the significance of the
developed technique and to discuss prospects for an actual application of the
formalism presented here to the description of hadrons---in particular, of
mesons---as bound states of quarks within the quantum field theory of the
strong interactions, quantum chromodynamics. A quantitative discussion of
these problems is certainly beyond the scope of the present approach.
Consequently, we'll content ourselves with the addition of a few modest and
merely qualitative comments.\begin{itemize}\item In order to justify at all
the consideration of the instantaneous approximation~to the Bethe--Salpeter
formalism, one should be able to answer the question: Which physical systems
may be reliably described by the instantaneous Bethe--Salpeter equation? In
other words, for which physical systems is the neglect of retardation effects
meaningful? Any sound quantitative statement about this problem would require
a comparison of the predictions for the bound-state masses of, on the one
hand, the full Bethe--Salpeter equation and, on the other hand, the
instantaneous Bethe--Salpeter equation. (This analysis would be similar to
the one performed~in Ref.~\cite{Olsson96} for the region of validity of the
reduced Salpeter equation.) However,~in order to carry out a study of this
kind, one must have at one's disposal a reliable method for extracting
information from the full Bethe--Salpeter equation; this~is, at present,
unfortunately not the case.\item In the hopefully pedagogical example chosen
for the introduction of our method, we confined ourselves to the (somewhat
simpler) case of bound-state constituents with vanishing masses. Clearly, the
next step in this game has to be the extension of this method to the case of
bound-state constituents with nonvanishing masses. The generalization to the
massive case is more or less straightforward (and can be done for an
arbitrary Lorentz structure of the Bethe--Salpeter interaction~kernel). Of
course, the nonvanishing values of the bound-state constituents' masses
entail a more complicated structure of the Bethe--Salpeter equation and,
consequently, a higher degree of complexity of all resulting expressions.
From our point of~view, this fact renders the results for the general case
not suitable for the introductory presentation of the main ideas.
Accordingly, the results generalizing this study~to massive bound-state
constituents have been reserved for a separate publication, see
Ref.~\cite{Lucha00:IBSEnzm}; our conclusions there are that the formalism
developed here may~be applied with due modifications to the same class of
potentials as considered here.\item One long-standing problem in meson
phenomenology is related to the question~of the Lorentz structure of the
confining Bethe--Salpeter interaction kernel. For the two reasons explained
in the Introduction---namely, stability of the solutions and linearity of the
mesonic Regge trajectories---we decided to focus our attention~to a confining
interaction kernel of time-component Lorentz-vector type.~However, as is now
well known \cite{lagae92II,muenz94,olsson95}, a time-component Lorentz-vector
confining kernel leads to a (from the phenomenological point of view) wrong
sign of the spin--orbit coupling and can thus not be considered a realistic
model for the confinement~of the colour degrees of freedom present in quantum
chromodynamics. Needless to say, our formalism can accommodate an arbitrary
Dirac structure of the kernel. For the most popular choices of the Lorentz
nature of the Bethe--Salpeter kernel, in particular, for an interaction
kernel of Lorentz-scalar ($1\otimes 1$), time-component Lorentz-vector
($\gamma^0\otimes\gamma^0$), and full Lorentz-vector
($\gamma^\mu\otimes\gamma_\mu$) type, the explicit form of the instantaneous
Bethe--Salpeter equation may be read off from Refs.~\cite{lagae92I,olsson95}.
\item Any realistic quark model for light mesons requires the inclusion of a
mechanism for the spontaneous breaking of chiral symmetry. Spontaneous chiral
symmetry breaking will manifest itself in the structure of the quark
propagators; the latter will deviate in their form from the form relevant for
the free case. In principle,~in our formalism the inclusion of spontaneous
chiral symmetry breaking in this~way should always be possible by application
of the appropriate expansions in terms of the basis states to the expressions
involving the (non-free) quark form factors.\item A proper embedding of
spontaneous chiral symmetry breaking should manage~to reconcile completely
the Bethe--Salpeter formalism with the Nambu--Goldstone theorem, which
demands the existence of a massless Nambu--Goldstone boson~for every
spontaneously broken global symmetry of the theory. This would allow~for the
interpretation of all light pseudoscalar mesons $\pi,$ K, $\eta$ as (pseudo-)
Goldstone bosons in the Bethe--Salpeter framework. (For a time-component
Lorentz-vector Bethe--Salpeter kernel, this mechanism has been shown to
operate in the desired way in Ref.~\cite{lagae92II}. This proves that the
existence of massive bound states of massless constituents is definitely no
inherent feature of the Bethe--Salpeter formalism.)\end{itemize}Thus, we may
be optimistic that our technique for solving the Bethe--Salpeter equation in
the instantaneous approximation by explicit construction of a matrix
representation of this eigenvalue equation will eventually become an
important tool for the description of hadrons as bound states of quarks from
first principles, quantum chromodynamics.

\section*{Acknowledgements}One of us (W.~L.) would like to thank
Michael~Beyer for encouraging discussions at~the ``International Symposium on
`Quarks in Hadrons and Nuclei'\,'' at Oberw\"olz,~Austria. Moreover, we would
like to thank Martin Olsson for several discussions of the results~of the
recent study of the instantaneous Bethe--Salpeter equation by him and
coworkers. One of us (K.~M.~M.) would like to thank the Erwin Schr\"odinger
International Institute for Mathematical Physics, where part of this work was
done, for hospitality and would like to acknowledge also the support by the
NSF under grant no.~HRD-9633750.

\newpage\appendix\section{The ``Generalized Laguerre''
Basis}\label{App:Laguerre}Our choice of the basis states $|\phi_i\rangle$ is
defined by their configuration-space representation
\begin{equation}\phi_i^{(\ell)}(r)=\sqrt{\frac{(2\,\mu)^{2\,\ell+3}\,i!}
{\Gamma(2\,\ell+i+3)}}\,r^\ell\exp(-\mu\,r)\, L_i^{(2\,\ell+2)}(2\,\mu\,r)\
,\quad i=0,1,2,\dots\ ,\label{Eq:IBSE-basis-config}\end{equation}which
involves the generalized Laguerre polynomials $L_i^{(\gamma)}(x)$ (for the
parameter $\gamma$):~the latter quantities are orthogonal polynomials which
are defined by the power series \cite{Abramowitz}$$L_i^{(\gamma)}(x)=
\sum_{t=0}^i\,(-1)^t\left(\begin{array}{c}i+\gamma\\i-t\end{array}\right)
\frac{x^t}{t!}\ ,\quad i=0,1,2,\dots\ ,$$and which are orthonormalized, with
the weight function $x^\gamma\exp(-x)$, according to~\cite{Abramowitz}
$$\int\limits_0^\infty{\rm d}x\,x^\gamma\exp(-x)\,L_i^{(\gamma)}(x)\,
L_j^{(\gamma)}(x)=\frac{\Gamma(\gamma+i+1)}{i!}\,\delta_{ij}\ ,\quad
i,j=0,1,2,\dots\ .$$The requirement of normalizability of the Hilbert-space
basis states $|\phi_i\rangle$ imposes on~the variational parameter $\mu$ the
constraint $$\mu>0\ .$$In this case, the configuration-space basis functions
$\phi_i^{(\ell)}(r),$ defined by Eq.~(\ref{Eq:IBSE-basis-config}), satisfy
the orthonormalization condition$$\int\limits_0^\infty{\rm d}r\,r^2\,
\phi_i^{(\ell)}(r)\,\phi_j^{(\ell)}(r)=\delta_{ij}\ ,\quad i,j=0,1,2,\dots\
.$$Note: the configuration-space representation of our basis states
$|\phi_i\rangle$ is chosen to be~real. The momentum-space representation of
these basis states, $\phi_i^{(\ell)}(p),$ obtained by Fourier transformation
of Eq.~(\ref{Eq:IBSE-basis-config}), reads\begin{eqnarray}\phi_i^{(\ell)}(p)
&=&\sqrt{\frac{(2\,\mu)^{2\,\ell+3}\,i!}{\Gamma(2\,\ell+i+3)}}\,\frac{(-{\rm
i})^\ell\,p^\ell}{2^{\ell+1/2}\,\Gamma\left(\ell+\frac{3}{2}\right)}\,
\nonumber\\[1ex]
&\times&\sum_{t=0}^i\,\frac{(-1)^t}{t!}\left(\begin{array}{c}i+2\,\ell+2\\
i-t\end{array}\right)\frac{\Gamma(2\,\ell+t+3)\,(2\,\mu)^t}
{(p^2+\mu^2)^{(2\,\ell+t+3)/2}}\nonumber\\[1ex]
&\times&F\left(\frac{2\,\ell+t+3}{2},-\frac{1+t}{2};\ell+\frac{3}{2};
\frac{p^2}{p^2+\mu^2}\right),\quad i=0,1,2,\dots\ ,\label{Eq:IBSE-basis-mom}
\end{eqnarray}with the hypergeometric series $F$, defined, in terms of the
gamma function $\Gamma$, by \cite{Abramowitz}
$$F(u,v;w;z)=\frac{\Gamma(w)}{\Gamma(u)\,\Gamma(v)}\,\sum_{n=0}^\infty\,
\frac{\Gamma(u+n)\,\Gamma(v+n)}{\Gamma(w+n)}\,\frac{z^n}{n!}\ .$$The
momentum-space basis functions $\phi_i^{(\ell)}(p)$ satisfy the
orthonormalization condition\begin{equation}\int\limits_0^\infty{\rm
d}p\,p^2\,\phi_i^{\ast(\ell)}(p)\,\phi_j^{(\ell)}(p)=\delta_{ij}\ ,\quad
i,j=0,1,2,\dots\ .\label{Eq:orthonorm-mom}\end{equation}The availability of
the Fourier transform of our basis functions $\phi_i^{(\ell)}(r)$ in analytic
form represents the main advantage of our choice
(\ref{Eq:IBSE-basis-config}). Note that the momentum-space basis functions
are real for $\ell=0,$ as well as for all even values of
$\ell$:$$\phi_i^{\ast(\ell)}(p)=\phi_i^{(\ell)}(p)\quad\mbox{for\
}\ell=0,2,4,\dots\ .$$

\section{Evaluation of the Matrix ${\cal M}_{ij}$}\label{App:conversion}
\subsection{Preliminaries}In the course of evaluating the various matrices
$A_{ij},\dots,D_{ij}$ summing up to the matrix ${\cal M}_{ij}$ according to
Eq.~(\ref{Eq:IBSE-matrix}), we occasionally encounter (``moment'') integrals
of the type\begin{equation}I^{(n)}_{ij}(\mu)\equiv\int\limits_0^\infty{\rm
d}k\,k^{2+n}\,\phi_i^{(0)}(k)\,\phi_j^{(0)}(k)\ ,\quad n=0,1,2,\dots\
.\label{Eq:nth-moment}\end{equation}According to the above definition, all
the matrices $I^{(n)}_{ij}(\mu)$ are both real and symmetric under
permutation of the indices $i$ and $j.$ Taking advantage of a somewhat
simplified expression for our momentum-space basis functions
$\phi_i^{(\ell)}(p)$ valid in the case $\ell=0$ \cite{Lucha97}, viz.,
\begin{eqnarray*}\phi_i^{(0)}(p)&=&\sqrt{\frac{i!}
{\mu\,\pi\,\Gamma(i+3)}}\,\frac{4}{p}\,\sum_{t=0}^i\,(-2)^t\,(t+1)
\left(\begin{array}{c}i+2\\i-t\end{array}\right)\\[1ex]
&\times&\left(1+\frac{p^2}{\mu^2}\right)^{-(t+2)/2}
\sin\left((t+2)\arctan\frac{p}{\mu}\right),\end{eqnarray*}it is rather
straightforward to find the (exact) analytic expression for the
integrals~(\ref{Eq:nth-moment}):
\begin{eqnarray}&&I^{(n)}_{ij}(\mu)\label{Eq:nth-moment-result}\\[1ex]&&=
\frac{4\,\mu^n}{\pi\,\sqrt{(i+1)\,(i+2)\,(j+1)\,(j+2)}}\nonumber\\[1ex]
&&\times\sum_{r=0}^i\,\sum_{s=0}^j\,(-2)^{r+s}
\left(\begin{array}{c}i+2\\i-r\end{array}\right)
\left(\begin{array}{c}j+2\\j-s\end{array}\right)(r+1)\,(s+1)
\nonumber\\[1ex]&&\times
\left[\sum_{k=0}^{|r-s|}\left(\begin{array}{c}|r-s|\\k\end{array}\right)
\frac{\Gamma(\frac{1}{2}\,(k+n+1))\,\Gamma(\frac{1}{2}\,(r+s+3+|r-s|-n-k))}
{\Gamma(\frac{1}{2}\,(r+s+4+|r-s|))}
\cos\left(\frac{k\,\pi}{2}\right)\right.\nonumber\\[1ex]
&&\left.-\sum_{k=0}^{r+s+4}\left(\begin{array}{c}r+s+4\\k\end{array}\right)
\frac{\Gamma(\frac{1}{2}\,(k+n+1))\,\Gamma(\frac{1}{2}\,(2\,r+2\,s+7-n-k))}
{\Gamma(r+s+4)}\cos\left(\frac{k\,\pi}{2}\right)\right].\nonumber\end{eqnarray}

Furthermore, we will need the (exact) analytic expressions of integrals of
the kind\begin{equation}J^{(n)}_{ij}(\mu)\equiv\int\limits_0^\infty{\rm
d}k\,k^{2+n}\,\phi_i^{\ast(1)}(k)\,\phi_j^{(0)}(k)\ ,\quad n=0,1,2,\dots\
.\label{Eq:nth-overlap}\end{equation}In order to get rid of the
(difficult-to-handle) hypergeometric series $F$ entering in the
momentum-space basis functions $\phi_i^{(1)}(p),$ we employ a suitable
recursion formula~\cite{Abramowitz}:\begin{eqnarray*}\phi_i^{(1)}(p)&=&-{\rm
i}\,\sqrt{\frac{\mu^5}{\pi\,(i+1)\,(i+2)\,(i+3)\,(i+4)}}
\,\frac{8}{p^2}\\[1ex]&\times&\sum_{t=0}^i\,\frac{(-2)^t}{t!}
\left(\begin{array}{c}i+4\\i-t\end{array}\right)
\frac{(t+3)!\,\mu^t}{(p^2+\mu^2)^{(t+3)/2}}\\[1ex]&\times&
\left[\frac{\sqrt{p^2+\mu^2}}{t+2}\sin\left((t+2)\arctan\frac{p}{\mu}\right)-
\frac{\mu}{t+3}\sin\left((t+3)\arctan\frac{p}{\mu}\right)\right].
\end{eqnarray*}Note: since $\phi_i^{(1)}(p)$ is purely imaginary, all these
matrices $J^{(n)}_{ij}(\mu)$ are purely imaginary. Again it is a tedious but
straightforward task to work out the integrals~(\ref{Eq:nth-overlap})
explicitly:\begin{eqnarray}&&J^{(n)}_{ij}(\mu)
\label{Eq:nth-overlap-result}\\[1ex]&&={\rm i}\,\frac{8\,\mu^n}
{\pi\,\sqrt{(i+1)\,(i+2)\,(i+3)\,(i+4)\,(j+1)\,(j+2)}}\nonumber\\[1ex]
&&\times\sum_{r=0}^i\,\sum_{s=0}^j\,(-2)^{r+s}\,(r+1)\,(r+2)\,(r+3)\,(s+1)
\left(\begin{array}{c}i+4\\i-r\end{array}\right)
\left(\begin{array}{c}j+2\\j-s\end{array}\right)\nonumber\\[1ex]
&&\times\left\{\frac{1}{r+2}\left[\sum_{k=0}^{|r-s|}
\left(\begin{array}{c}|r-s|\\k\end{array}\right)
\frac{\Gamma(\frac{1}{2}\,(n+k))\,
\Gamma(\frac{1}{2}\,(4+r+s+|r-s|-n-k))}
{\Gamma(\frac{1}{2}\,(4+r+s+|r-s|))}\right.\right.\nonumber\\[1ex]
&&\times\cos\left(\frac{k\,\pi}{2}\right)\nonumber\\[1ex]
&&-\left.\sum_{k=0}^{4+r+s}\,\left(\begin{array}{c}4+r+s\\k\end{array}\right)
\frac{\Gamma(\frac{1}{2}\,(n+k))\,\Gamma(\frac{1}{2}\,(8+2\,r+2\,s-n-k))}
{\Gamma(4+r+s)}\cos\left(\frac{k\,\pi}{2}\right)\right]\nonumber\\[1ex]
&&-\frac{1}{r+3}\left[\sum_{k=0}^{|1+r-s|}\left(\begin{array}{c}
|1+r-s|\\k\end{array}\right)\right.\nonumber\\[1ex]&&\times
\frac{\Gamma(\frac{1}{2}\,(n+k))\,\Gamma(\frac{1}{2}\,(5+r+s+|1+r-s|-n-k))}
{\Gamma(\frac{1}{2}\,(5+r+s+|1+r-s|))}\cos\left(\frac{k\,\pi}{2}\right)
\nonumber\\[1ex]&&\left.\left.-\sum_{k=0}^{5+r+s}
\left(\begin{array}{c}5+r+s\\k\end{array}\right)
\frac{\Gamma(\frac{1}{2}\,(n+k))\,\Gamma(\frac{1}{2}\,(10+2\,r+2\,s-n-k))}
{\Gamma(5+r+s)}\cos\left(\frac{k\,\pi}{2}\right)\right]\right\}.\nonumber
\end{eqnarray}

The radial basis functions $\phi_i^{(\ell)}(r)$ and $\phi_i^{(\ell)}(p)$ in
configuration and momentum space given in Eqs.~(\ref{Eq:IBSE-basis-config})
and (\ref{Eq:IBSE-basis-mom}), respectively, are derived from the respective
basis functions on the (full) three-dimensional Euclidean space $R^3$ by
factorizing off the dependence on the angular variables. This dependence is
described by the spherical harmonics~${\cal Y}_{\ell m}(\Omega)$ for angular
momentum $\ell=0,1,2,\dots$ and its projection $m=-\ell,-\ell+1,\dots,+\ell,$
which depend on the solid angle $\Omega$ and are orthonormalized according
to$$\int{\rm d}\Omega\,{\cal Y}^\ast_{\ell m}(\Omega)\,{\cal
Y}_{\ell'm'}(\Omega)=\delta_{\ell\ell'}\,\delta_{mm'}\ .$$Since the two sets
of basis functions on $R^3$ are related by a Fourier transformation,~the
corresponding radial basis functions $\phi_i^{(\ell)}(r)$ and
$\phi_i^{(\ell)}(p),$ $i=0,1,2,\dots,$ are related by\begin{eqnarray}
\phi_i^{(\ell)}(r)&=&{\rm
i}^\ell\,\sqrt{\frac{2}{\pi}}\int\limits_0^\infty{\rm
d}p\,p^2\,j_\ell(p\,r)\,\phi_i^{(\ell)}(p)\
,\label{Eq:Fourier-Bessel-config}\\[1ex]\phi_i^{(\ell)}(p)&=&(-{\rm
i})^\ell\,\sqrt{\frac{2}{\pi}}\int\limits_0^\infty{\rm
d}r\,r^2\,j_\ell(p\,r)\,\phi_i^{(\ell)}(r)\
,\label{Eq:Fourier-Bessel-mom}\end{eqnarray}which may be easily seen with the
help of the well-known expansion of the plane waves in terms of spherical
harmonics ${\cal Y}_{\ell m}$ in configuration $(\Omega_{\bf x})$ and
momentum $(\Omega_{\bf p})$ space,$$\exp({\rm i}\,{\bf p}\cdot{\bf
x})=4\pi\,\sum_{\ell=0}^\infty\,\sum_{m=-\ell}^{+\ell}\,{\rm
i}^\ell\,j_\ell(p\,r)\,{\cal Y}^\ast_{\ell m}(\Omega_{\bf p})\,{\cal Y}_{\ell
m}(\Omega_{\bf x})\ .$$\subsection{Evaluation of $A_{ij}$}Expressed in terms
of the integral $I^{(n)}_{ij}(\mu)$ introduced in Eq.~(\ref{Eq:nth-moment}),
the term $A_{ij}$ in the matrix ${\cal M}_{ij},$ defined in
Eq.~(\ref{Eq:IBSE-term-I}), is simply given by$$A_{ij}=4\,I^{(2)}_{ij}(\mu)\
.$$\subsection{Evaluation of $B_{ij}$}The term $B_{ij}$ in the matrix ${\cal
M}_{ij},$ Eq.~(\ref{Eq:IBSE-term-II}), becomes, for some interaction
potential~$V(r)$, which enters in this term linearly in form of
Eq.~(\ref{Eq:IBSE-intpot}), $$B_{ij}=\frac{4}{\pi}\int\limits_0^\infty{\rm
d}r\,r^2\,V(r)\int\limits_0^\infty{\rm d}k\,k^3\,\phi_i^{(0)}(k)\,j_0(k\,r)
\int\limits_0^\infty{\rm d}k'\,k'^2\,j_0(k'\,r)\,\phi_j^{(0)}(k')\ .$$In
order to be able to apply the Fourier--Bessel relation
(\ref{Eq:Fourier-Bessel-config}), we expand the expression
$k\,\phi_i^{(0)}(k)$ in terms of the momentum-space basis functions
$\phi_i^{(0)}(k)$:
\begin{equation}k\,\phi_i^{(0)}(k)=\sum_{j=0}^N\,b_{ji}\,\phi_j^{(0)}(k)\
.\label{Eq:expansion-b}\end{equation}As a consequence of the
orthonormality~(\ref{Eq:orthonorm-mom}) of the (momentum-space) basis
functions $\phi_i^{(\ell)}(p),$ the expansion coefficients $b_{ij}$ may be
expressed in terms of the integral $I^{(1)}_{ij}(\mu)$ defined in
Eq.~(\ref{Eq:nth-moment}):\begin{equation}b_{ij}=\int\limits_0^\infty{\rm
d}k\,k^3\,\phi_i^{(0)}(k)\,\phi_j^{(0)}(k)\equiv I^{(1)}_{ij}(\mu)\
.\label{Eq:b=I1}\end{equation}Inserting the expansion (\ref{Eq:expansion-b})
involving the coefficients $b_{ij}$, applying the Fourier--Bessel relation
(\ref{Eq:Fourier-Bessel-config}), and remembering the definition
(\ref{Eq:expval-intpot}) of the expectation values $V^{(\ell)}_{ij}$ of the
interaction potential $V(r)$ under consideration, the term $B_{ij}$ reduces
to the---at least, for all power-law potentials algebraic---expression
\begin{equation}B_{ij}=2\,\sum_{r=0}^N\,b_{ri}(\mu)\,V^{(0)}_{rj}(\mu)\
.\label{Eq:term-II-expansion}\end{equation}\subsection{Evaluation of
$C_{ij}$}The term $C_{ij}$ in the matrix ${\cal M}_{ij},$
Eq.~(\ref{Eq:IBSE-term-III}), becomes, for some interaction potential~$V(r)$,
which enters in this term linearly in form of
Eq.~(\ref{Eq:IBSE-intpot}),$$C_{ij}=\frac{4}{\pi}\int\limits_0^\infty{\rm
d}r\,r^2\,V(r)\int\limits_0^\infty{\rm d}k\,k^2\,\phi_i^{(0)}(k)\,j_1(k\,r)
\int\limits_0^\infty{\rm d}k'\,k'^3\,j_1(k'\,r)\,\phi_j^{(0)}(k')\ .$$In
order to be able to apply the Fourier--Bessel relation
(\ref{Eq:Fourier-Bessel-config}), we are forced to expand both the
momentum-space basis functions $\phi_i^{(0)}(k)$ and the expression
$k\,\phi_i^{(0)}(k)$ in terms of the momentum-space basis functions
$\phi_i^{(1)}(k)$:\begin{eqnarray}\phi_i^{(0)}(k)&=&
\sum_{j=0}^N\,c_{ji}\,\phi_j^{(1)}(k)\ ,\label{Eq:expansion-c}\\[1ex]
k\,\phi_i^{(0)}(k)&=&\sum_{j=0}^N\,d_{ji}\,\phi_j^{(1)}(k)\
.\label{Eq:expansion-d}\end{eqnarray}As a consequence of the
orthonormality~(\ref{Eq:orthonorm-mom}) of the (momentum-space) basis
functions $\phi_i^{(\ell)}(p),$ the expansion coefficients $c_{ij}$ and
$d_{ij}$ may be expressed in terms of the integrals $J^{(n)}_{ij}(\mu),$
$n=0,1,$ defined in Eq.~(\ref{Eq:nth-overlap}):\begin{eqnarray}c_{ij}&=&
\int\limits_0^\infty{\rm d}k\,k^2\,\phi_i^{\ast(1)}(k)\,\phi_j^{(0)}(k)\equiv
J^{(0)}_{ij}(\mu)\ ,\label{Eq:c=J0}\\[1ex]d_{ij}&=&\int\limits_0^\infty{\rm
d}k\,k^3\,\phi_i^{\ast(1)}(k)\,\phi_j^{(0)}(k)\equiv J^{(1)}_{ij}(\mu)\
.\label{Eq:d=J1}\end{eqnarray}Inserting the expansions (\ref{Eq:expansion-c})
and (\ref{Eq:expansion-d}) involving the coefficients $c_{ij}$ and $d_{ij}$,
applying the Fourier--Bessel relation (\ref{Eq:Fourier-Bessel-config}), and
remembering the definition (\ref{Eq:expval-intpot}) of the expectation values
$V^{(\ell)}_{ij}$ of the interaction potential $V(r)$ under consideration,
the term $C_{ij}$ reduces to the---at least, for all power-law potentials
algebraic---expression\begin{equation}C_{ij}=2\,\sum_{r=0}^N\,\sum_{s=0}^N\,
c^\ast_{ri}\,d_{sj}(\mu)\,V^{(1)}_{rs}(\mu)\
.\label{Eq:term-III-expansion}\end{equation}\subsection{Evaluation of
$D_{ij}$}The term $D_{ij}$ in the matrix ${\cal M}_{ij},$
Eq.~(\ref{Eq:IBSE-term-IV}), becomes, for some interaction potential~$V(r)$,
which enters in this term quadratically in form of
Eq.~(\ref{Eq:IBSE-intpot}),
\begin{eqnarray*}D_{ij}&=&\frac{4}{\pi^2}
\int\limits_0^\infty{\rm d}r\,r^2\,V(r)
\int\limits_0^\infty{\rm d}r'\,r'^2\,V(r')
\int\limits_0^\infty{\rm d}k\,k^2\,\phi_i^{(0)}(k)\,j_1(k\,r)\\[1ex]&\times&
\int\limits_0^\infty{\rm d}k'\,k'^2\,j_1(k'\,r)\,j_0(k'\,r')
\int\limits_0^\infty{\rm d}k''\,k''^2\,j_0(k''\,r')\,\phi_j^{(0)}(k'')\
.\end{eqnarray*}Adopting the expansion (\ref{Eq:expansion-c}) of
$\phi_i^{(0)}(k)$ in terms of $\phi_i^{(1)}(k)$---which introduces again~the
coefficients $c_{ij}$---it is clearly no problem to apply twice the
Fourier--Bessel relation (\ref{Eq:Fourier-Bessel-config}):$$D_{ij}={\rm
i}\,\frac{2}{\pi}\,\sum_{s=0}^N\,c^\ast_{si}
\int\limits_0^\infty{\rm d}r\,r^2\,V(r)\,\phi_s^{(1)}(r)
\int\limits_0^\infty{\rm d}r'\,r'^2\,V(r')\,\phi_j^{(0)}(r')
\int\limits_0^\infty{\rm d}k\,k^2\,j_1(k\,r)\,j_0(k\,r')\ .$$However, in
order to be able to apply the Fourier--Bessel relation
(\ref{Eq:Fourier-Bessel-mom}), we are forced~to expand the expressions
$V(r)\,\phi_i^{(\ell)}(r),$ $\ell=0,1,$ over the appropriate
configuration-space basis functions $\phi_i^{(\ell)}(r),$ $\ell=0,1,$
respectively. The corresponding expansion coefficients are the expectation
values $V^{(\ell)}_{ij}(\mu)$ of the interaction potential $V(r),$ defined by
Eq.~(\ref{Eq:expval-intpot}):\begin{equation}V(r)\,\phi_i^{(\ell)}(r)=
\sum_{j=0}^N\,V^{(\ell)}_{ji}(\mu)\,\phi_j^{(\ell)}(r)\ ,\quad\ell=0,1\
.\label{Eq:expansion-V}\end{equation}Inserting these expansions involving the
expectation values $V^{(\ell)}_{ij}(\mu)$ of the interaction potential $V(r)$
in Eq.~(\ref{Eq:expval-intpot}), applying the Fourier--Bessel relation
(\ref{Eq:Fourier-Bessel-mom}), and remembering the orthonormalization
relation (\ref{Eq:orthonorm-mom}) satisfied by the momentum-space basis
functions $\phi_i^{(\ell)}(p),$ the term $D_{ij}$ reduces to the---at least,
for all power-law potentials algebraic---expression$$D_{ij}=
\sum_{r=0}^N\,\sum_{s=0}^N\,\sum_{t=0}^N\,c^\ast_{ri}\,c_{st}\,
V^{(1)}_{sr}(\mu)\,V^{(0)}_{tj}(\mu)\ .$$\subsection{Relations Among the
Expansion Coefficients}The expansion coefficients $b_{ij},$ $c_{ij},$
$d_{ij}$ defined by Eqs.~(\ref{Eq:expansion-b}), (\ref{Eq:expansion-c}),
(\ref{Eq:expansion-d}), respectively,~are, of course, not independent; they
satisfy the (only in the limit $N\to\infty,$ exact)
relations\begin{eqnarray}&&\sum_{r=0}^N\,c^\ast_{ri}\,c_{rj}=\delta_{ij}\ ,
\label{Eq:c*c}\\[1ex]&&\sum_{r=0}^N\,c^\ast_{ri}\,d_{rj}(\mu)=\sum_{r=0}^N\,
d^\ast_{ri}(\mu)\,c_{rj}=I^{(1)}_{ij}(\mu)\equiv b_{ij}(\mu)\ ,\label{Eq:c*d}
\\[1ex]&&\sum_{r=0}^N\,d^\ast_{ri}(\mu)\,d_{rj}(\mu)=I^{(2)}_{ij}(\mu)\
.\label{Eq:d*d}\end{eqnarray}Clearly, relation~(\ref{Eq:c*c}) expresses only
the unitarity (in the limit $N\to\infty$) of the matrix $c\equiv(c_{ij})$
which represents, according to its definition~(\ref{Eq:expansion-c}), the
transformation from~the basis $\{\phi_i^{(1)}(p),i=0,1,2,\dots\}$ to the
basis $\{\phi_i^{(0)}(p),i=0,1,2,\dots\}$ of the space
$L_2(R^+).$\subsection{Accuracy Considerations}\label{Sec:Accuracy}The
existence of relations like the ones in Eqs.~(\ref{Eq:c*c}), (\ref{Eq:c*d}),
and (\ref{Eq:d*d}) opens a possibility~to investigate systematically the
errors induced by the truncations of the expansion series in the terms
$B_{ij},$ $C_{ij},$ and $D_{ij}.$ For instance, for $d=15$ (i.e., $15\times
15$ matrices) and~for $N=49$ (i.e., a truncation to the first 50 basis
states), the relative error is less than~$3\%.$

Furthermore, taking advantage of the fact that the momentum-space basis
function $\phi_0^{(0)}(p)$ may be cast into the simple form$$\phi_0^{(0)}(p)=
4\,\sqrt{\frac{2}{\pi}}\,\frac{\mu^{5/2}}{(p^2+\mu^2)^2}\ ,$$the first
entries of the terms $B_{ij}$ and $C_{ij}$ defined in
Eqs.~(\ref{Eq:IBSE-term-II}) and (\ref{Eq:IBSE-term-III}),
respectively,~can~be evaluated, for appropriate choices of the interaction
potential $V(r),$ analytically. These explicit results may be compared with
the corresponding approximate values obtained from the series expansions
(\ref{Eq:term-II-expansion}) and (\ref{Eq:term-III-expansion}), respectively.
For the linear potential (\ref{Eq:linpot}),~the numerical value of the term
$B_{00}$ is found to be$$\frac{4\pi}{\lambda}\,B_{00}\equiv
\frac{2}{\pi\,\lambda}\int\limits_0^\infty{\rm d}k\,k^3\,\phi_0^{(0)}(k)
\int\limits_0^\infty{\rm d}k'\,k'^2\,V_0(k,k')\,\phi_0^{(0)}(k')=\frac{64}{3}
=21.\dot 3\ .$$This has to be compared with the series expansion
$$\frac{8\pi}{\lambda}\,\sum_{r=0}^N\,b_{r0}(\mu)\,V^{(0)}_{r0}(\mu)=
\frac{8\pi}{\lambda}\,\sum_{r=0}^N\,b_{r0}(1)\,V^{(0)}_{r0}(1)\ ,$$which
yields, for, e.g., $N=49,$ that is, taking into account the first 50 basis
functions,$$\frac{8\pi}{\lambda}\,\sum_{r=0}^{49}\,b_{r0}(1)\,V^{(0)}_{r0}(1)
=21.333333333333334562\ .$$Similarly, for a constant potential, i.e.,
$V(r)=\lambda,$ the numerical value of the term $C_{00}$ is found to
be$$\frac{4\pi}{\lambda\,\mu}\,C_{00}\equiv\frac{2}{\pi\,\lambda\,\mu}
\int\limits_0^\infty{\rm d}k\,k^2\,\phi_0^{(0)}(k)
\int\limits_0^\infty{\rm d}k'\,k'^3\,V_1(k,k')\,\phi_0^{(0)}(k')=\frac{64}{3}
=21.\dot 3\ .$$This has to be compared with the series expansion
$$\frac{8\pi}{\lambda\,\mu}\,\sum_{r=0}^N\,\sum_{s=0}^N\,c^\ast_{r0}\,
d_{s0}(\mu)\,V^{(1)}_{rs}(\mu)=8\pi\,\sum_{r=0}^N\,c^\ast_{r0}\,d_{r0}(1)\
,$$which yields, for, e.g., $N=49,$ that is, taking into account the first 50
basis functions,$$8\pi\,\sum_{r=0}^{49}\,c^\ast_{r0}\,d_{r0}(1)=21.333307054\
.$$Beyond doubt, this accuracy of the series expansions should suffice for
our purposes.

\end{document}